\documentclass{kapproc} 

\usepackage{procps} 

\usepackage[dvips]{graphicx}

\upperandlowercase

\let\footnote\savefootnote

\normallatexbib 

\begin{document}

\articletitle[On the Intracluster Medium...]{On the Intracluster Medium in\\ 
Cooling Flow \& Non-Cooling Flow Clusters\footnote{Research supported by 
NSERC (Canada) Discovery Grant Award}
}

\author{Arif Babul, Ian G. McCarthy \& Greg B. Poole} 

\affil{Department of Physics and Astronomy\\
University of Victoria, 3800 Finnerty Road\\
Victoria, BC, Canada.  V8P 1A1}
\email{babul@uvic.ca}


\anxx{Babul\, Arif}
\anxx{McCarthy\, Ian G.}
\anxx{Poole\, Greg B.}

\begin{abstract}
Recent X-ray observations have highlighted clusters that lack entropy cores.  At first glance,
these results appear to invalidate the preheated ICM models.  We show that a self-consistent
preheating model, which factors in the effects of radiative cooling, is in excellent agreement
with the observations.  Moreover, the model naturally explains the intrinsic scatter in the 
L-T relation, with ``cooling flow'' and ``non-cooling flow'' systems corresponding to mildly 
and strongly preheated systems, respectively.  We discuss why preheating ought to be favoured 
over merging as a mechanism for the origin of ``non-cooling flow'' clusters.
\end{abstract}


\section{Introduction}
Correlations between the various X-ray and Sunyaev-Zel'dovich Effect (SZE)
properties of galaxy clusters offer important clues into the physical processes 
that have impacted the intracluster medium (ICM).   Observed scaling
relations have been shown to deviate significiantly from expectations
based on numerical simulations and analytic models that only take 
into account the influence of gravity on the ICM.
Such discrepancies have prompted considerations of additional, previously unexamined, 
gas physics.  One model, the preheating model, explores the possibility that
the nascent ICM is heated by galactic winds and/or AGN outflows
from galaxies existing at the time.  Even in its simplest avatar, the model scaling
relations, be they SZE v.~SZE, SZE v.~X-ray or X-ray v.~X-ray, are 
in remarkable agreement with the observations 
(cf.~\cite{Betal02, Metal02, Metal03}).  
However, recent X-ray data from \textit{XMM-Newton} and \textit{Chandra} 
suggests a potential problem with the model.  Preheating of the ICM
sets an entropy floor that manifests itself as a central core-like 
structure in the entropy profile.  A number of observed profiles show no such cores. 
Although we recognize that the current set of published {\textit Chandra} and 
{\textit XMM-Newton} cluster results are biased in favour of ``massive cooling flow'' systems
or active mergers,
the very existence of systems with power-law-like
entropy profiles needs addressing.

Here, we briefly report on our efforts to understand this particular issue,
and its implications for the preheated model.  
Our investigations involve a re-examination of the assumptions 
underlying the theoretical model as well as of the observational evidence for and against 
the model.

\section{Re-examining the Theoretical Model}
Most
preheating models are incomplete in that
they do not take into account radiative cooling (cf. \cite{Betal02}).  While preheating 
lowers the efficiency of cooling, it does not mitigate it entirely and over a Hubble time, 
the effects of cooling can be significant.  Several recent studies, \cite{ 
Vetal02, Detal02},
have highlighted the potentially important role of cooling though not necessarily in the context
of a preheated model.   Traditionally, cooling has been difficult to model. However,
we have developed a fast, efficient scheme for doing so.  The scheme can factor in
the effects of not only preheating and radiative cooling 
(due to both line and continuum emission) 
on the ICM over cosmological timescales, but potentially also those due to other 
processes such as conduction.  The scheme is currently being tested against detailed 
hydrodynamic simulations and the initial results are very encouraging.
A detailed description of this scheme will be forthcoming.  Here
we present some early results for the preheated+cooling model.  

\section{Cluster Entropy Profiles: Theory and Observations}

\begin{figure}[t]
  \begin{minipage}[t]{0.5\linewidth}
    \centering
    \includegraphics[width=6cm]{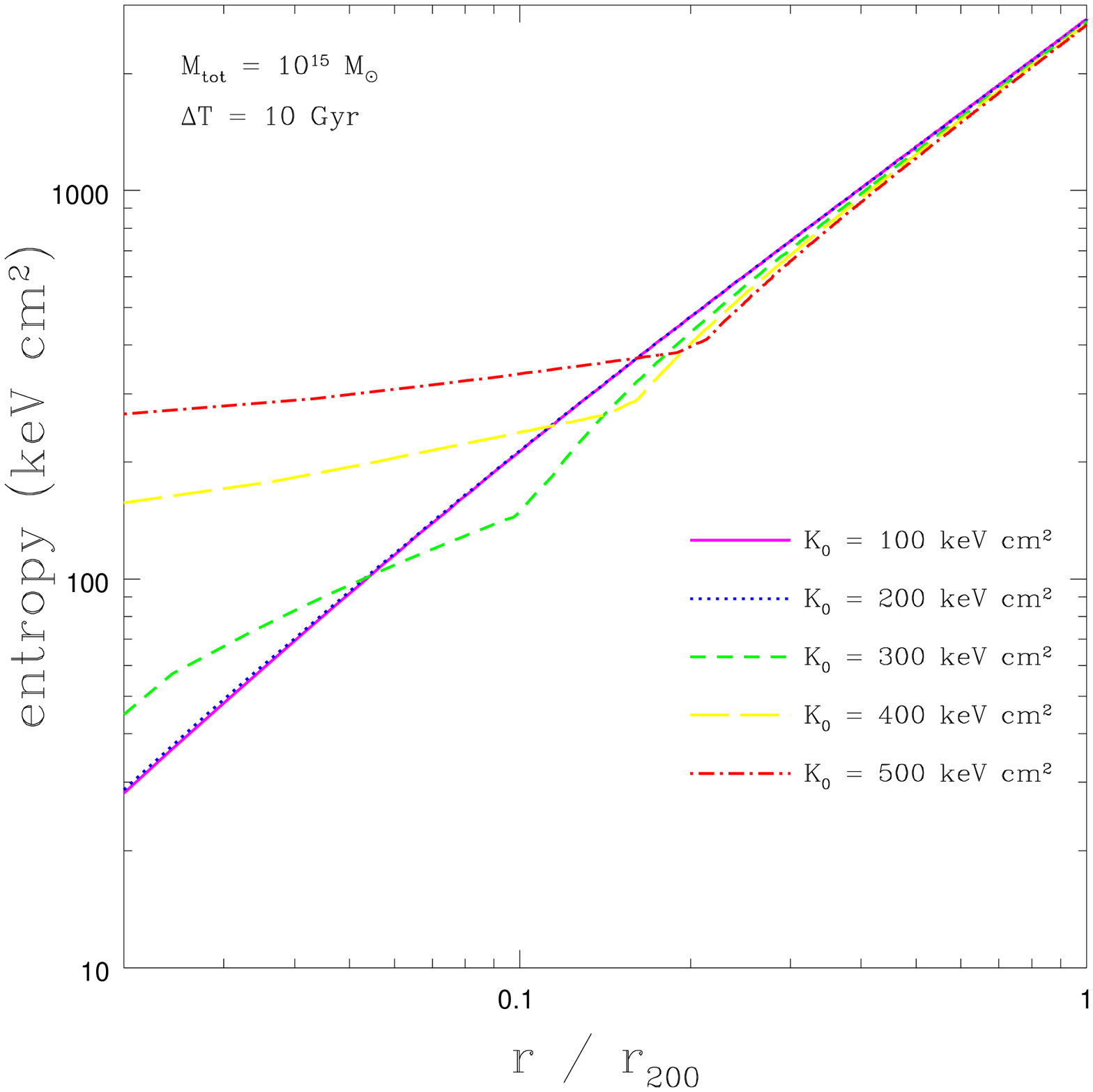}
\vskip-0.3cm
    \caption{Entropy profiles of a $10^{15}$M$_\odot$ cluster preheated to different values of 
    $K_\circ$ and observed after 10 Gyrs.  The efficacy of radiative cooling increases with decreasing
    $K_\circ$.  In clusters with low $K_\circ$ values, cooling rapidly erases the central core and 
    drives the profile into a $r^{1.1}$ power-law, reminiscent of the observed ``cooling flow'' clusters
    profiles.  Intriguingly, entropy profiles of actual clusters span the entire range of profile shapes shown.}
 \label{figure3}
  \end{minipage}%
\hspace{0.5cm} 
  \begin{minipage}[t]{0.5\linewidth}
    \centering
    \includegraphics[width=6cm]{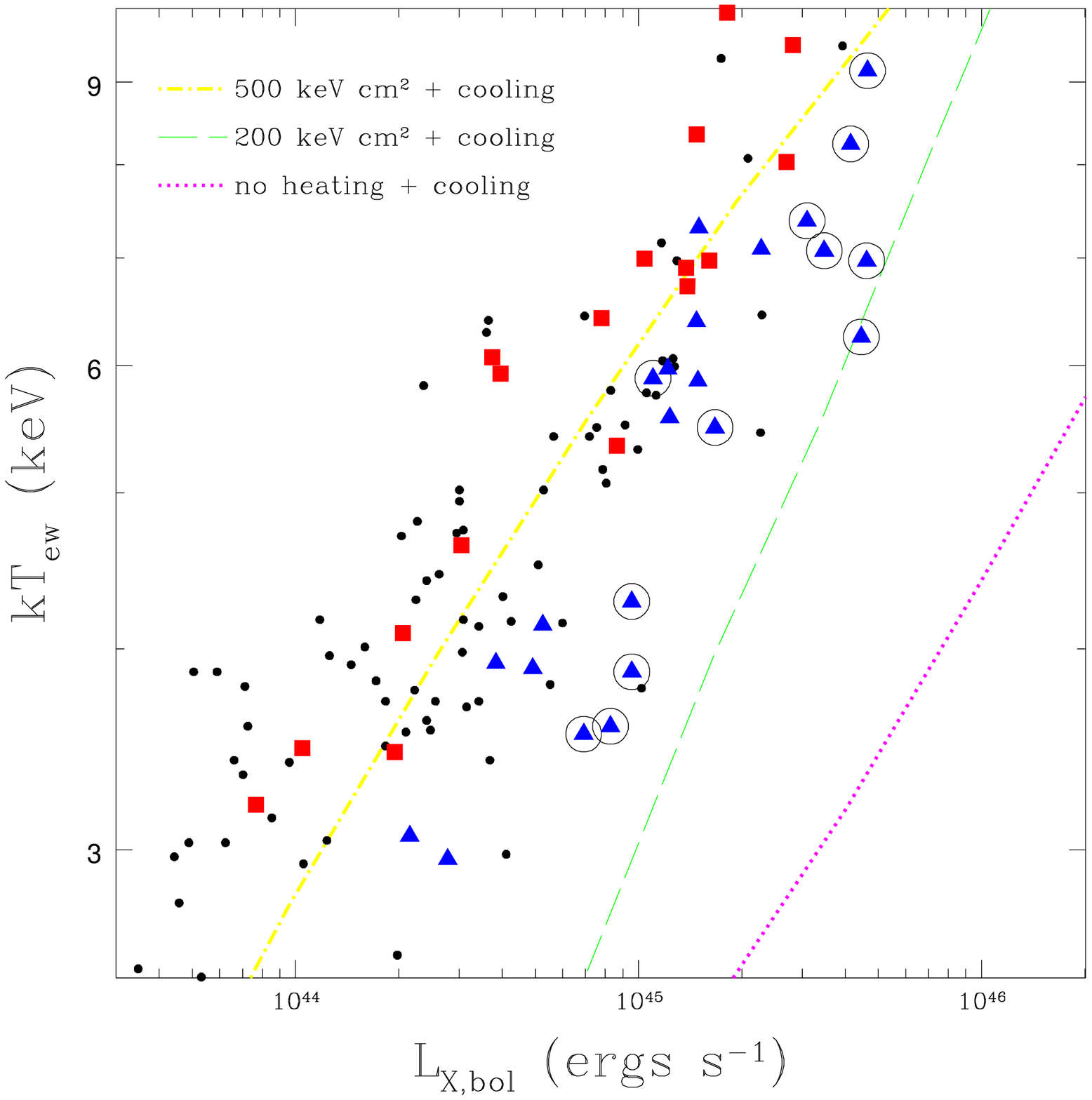}
\vskip-0.3cm
    \caption{The $L_X - T_X$ relation for $z<0.2$ clusters constructed using the 
actual observed values, as opposed to ``cooling flow 
corrected'' quantities. Within this locus, we identify the locations of 
known ``cooling flow'' and ``non-cooling flow'' clusters (see text for details).  
We also show the $L_X-T_X$ results for the preheated+cooling model. We identify NCF and
mCF systems with high and  low $K_\circ$ values, respectively.}
 \label{figure4}
  \end{minipage}%
\vskip-0.3cm
\end{figure}

Figure \ref{figure3} shows 
the effects of varying levels of preheating+cooling on cluster entropy profiles.
Radiative cooling is very efficient in clusters subjected to low levels of preheating.  
It causes the central entropy core to disappear on a relatively short timescale and 
drives the entropy profile into the $r^{1.1}$ power-law form reminiscent of the observed 
profiles of ``cooling flow'' (CF) clusters.  In contrast, cooling is much less 
efficient in clusters with highly preheated ICM.

We have assembled a preliminary collection of observed entropy profiles in order to see
how these compare to the model results.
We find, suprisingly, that the observed entropy profiles are not all self-similar and 
power-law-like, but span the entire range of shapes seen in Figure \ref{figure3}.  

\section{Reconstructing the Scaling Relations}

Figure \ref{figure4} shows $L_X-T_X$ relation for preheated+cooling clusters subjected to 
varying levels of preheating and observed $10$ Gyrs later.  
Also plotted are the 
($L_X$, $T_X$) for $z<0.2$ clusters  from Horner's ASCA Cluster Catalogue 
\cite{Hetal01}.  Unlike the data typically used in the construction of such plots, this set
is \textit{not} ``cooling flow corrected''.  This is important.
The theoretical values incorporate the effects of cooling; therefore, a fair comparison 
with the observations requires that we use data of comparable character.

Both the corrected and the uncorrected data exhibit the same correlations, though the latter
has larger scatter.  This is well known.
However, 
the scatter is not random.  Based 
on features in the temperature and X-ray SB profiles,  we have classified clusters as 
non-cooling flow clusters (NCF - squares), ordinary cooling flow clusters (CF - triangles), or massive cooling flow 
clusters (mCF - circled triangles).  Although most clusters remain unclassified, it is readily apparent that 
NCF systems lie close to the upper-left edge of the 
band while the mCF clusters delineate the opposite (lower-right) edge.

Comparing the preheated+cooling model predictions against the observations, we find the two 
in excellent agreement.  However, in order to account for the breadth and structure within the 
observed $L_X-T_X$ band, we have to abandon our previous \textit{ad hoc} assumption of uniform 
energy injection across the entire cluster
population.  Within our framework, the NCFs correspond to strongly preheated systems 
($K_\circ\sim 400$--$500$ Kev cm$^2$) while the mCFs correspond to mildly preheated systems 
($K_\circ\sim 100$--$200$ Kev cm$^2$).

\section{Non-Cooling Flow Clusters:  Products of Preheating?}

The assertion that NCF systems are strongly preheated clusters runs counter to the 
prevailing view.  In the latter, NCFs are identified as
clusters whose cool dense gas cores, the source of the excess central X-ray
emission characteristic of the CF clusters, have been disrupted by
major mergers.  Images of NCFs with disturbed X-ray morphologies are often used to 
support this scenario.  However, there are numerous CF systems
that also appear to be in the throes of on-going mergers.  Perseus is one such 
example \cite{C03}.  The ubiquity of mergers argues against them being the 
cause of the differences between CFs and NCFs. 

To test our hypothesis,
we have carried out a series of numerical simulation experiments.  
One distinguishing feature of our study is that our simulations include
radiative cooling.  Preliminary results suggests that even for nearly 
head-on 3:1 mergers, variations in the X-ray observables of the primary 
cluster are extremely short-lived.  In particular, we find that if the 
primary starts out as a CF-like system, by the time the merger remnant has been 
assimilated, it will have regained its CF-like character.  Motl et al. \cite{M03} 
too get similar results.  These findings argue that there ought \textit{not} to 
be any dynamically relaxed NCF systems.  But there are: eg.~A1413, A1651, A2319, A3158.
That such systems exist at all further indicates an alternate origin for the
CF/NCF clusters.

To reiterate, a self-consistent model of the ICM that factors in radiative processes
and allows for cluster-to-cluster variations in the level of initial preheating not only 
is able to account for the existence and the properties of CF and NCF clusters, but 
also of those that are between these two extremes.

\begin{acknowledgments}
We wish to thank our collaborators, M.~Balogh, G.~Holder, M.~Fardal, T.~Quinn, and
D.~Horner for their contributions to the work described here.
\end{acknowledgments}

\end{document}